\documentclass[english,a4paper,10pt]{article}

\usepackage[hyphens]{url}
\usepackage[natbibapa]{apacite}
\let\cite\citet

\usepackage{amssymb,amsmath,amsthm}
\usepackage{setspace}
\usepackage[shortlabels]{enumitem}

\usepackage[toc,page]{appendix}
\usepackage{subcaption,graphicx,float,adjustbox}
\usepackage{booktabs,tabularx,siunitx,threeparttable,dsfont,multirow}
\usepackage{tabularray}
\usepackage{pdflscape}
\usepackage{caption}
\usepackage{changes} % This pull in todonotes, too: \todo[inline]{Long comment}

\usepackage[T1]{fontenc}
\usepackage{mathpazo}
\usepackage{helvet}
\usepackage[english]{babel}

\usepackage[final,tracking=true,kerning=true,spacing=true,factor=1100,stretch=10,shrink=10]{microtype}
\microtypesetup{verbose=silent}
\usepackage{orcidlink}
\hypersetup{colorlinks=true, linkcolor=blue, citecolor=blue, urlcolor=blue, pdfborder={0 0 0}, bookmarks=true}
\usepackage{cleveref}

\theoremstyle{definition}

\newtheorem{hypothesis}{Hypothesis}

\newtheorem{result}{Result}

\microtypecontext{spacing=french}

\title{Paternalism and Deliberation: An Experiment on Making Formal Rules}
\author{%
  \large Max R. P. Grossmann\,\orcidlink{0000-0002-6152-9042}\\[2pt]
  {\footnotesize The University of Melbourne, Australia}\\
  {\footnotesize \href{mailto:GrossmannM@unimelb.edu.au}{GrossmannM@unimelb.edu.au}}
}
\date{\today}

\newcommand{\eg}[1]{\citep[e.g.,][]{#1}}

\begin{document}
\frenchspacing
\maketitle

\begin{abstract}
    Mandatory waiting periods have been instituted for medical procedures, gun purchases, and other high-stakes decisions. Are these softly paternalistic policies substitutes for harder restrictions, and are delayed decisions more respected?
In a general population survey experiment, Choice Architects make rules for decision-makers facing a high-stakes Bomb Risk Elicitation Task. Treatments vary when the decision is made: on the spot or after one day, and whether the initial decision can be revised. Choice Architects set a cap on the decision-maker's risk taking; in one treatment, they can additionally implement a mandatory waiting period.
Exogenous deliberation has no effect on the cap; equivalence testing (TOST) and Bayesian analysis ($\text{BF}_{01} \approx 12$) provide strong positive evidence for the absence of an effect. Endogenously prescribed waiting periods are add-on restrictions that do not substitute for the cap. Choice Architects believe that, with time, the average decision-maker will take less risk and---because of the distribution of Choice Architects' bliss points---come closer to Choice Architects' subjective ideal choice; the resulting reduction in forecasted errors is small.
Soft and hard paternalistic instruments are not substitutes: waiting periods are deployed as add-on restrictions.

    \par\vspace{1em}

    \noindent\textsc{JEL Classification}: C91, D04, D78, D91, I18, I31, I38\par\vspace{1em}

    \noindent\textsc{Keywords}: Paternalism, waiting periods, policy mix, high stakes, Bomb Risk Elicitation Task
\end{abstract}

\section*{Acknowledgments}

We thank David Albrecht, Arno Apffelstaedt, James Druckman, Hannah Hamilton, Felix Kölle, Melisa Kurtis, Aenne Läufer, Axel Ockenfels and Rastislav Rehák, audiences at Berlin, Buckingham, Cologne, Maastricht and Riga and two anonymous reviewers at Time-sharing Experiments for the Social Sciences (TESS) for valuable comments. We thank TESS and the NORC at the University of Chicago for outstanding survey-related support. We are grateful to Ben Mrakovcic for his implementation of the Chooser experiment. We acknowledge the use of artificial intelligence for the purposes of research assistance and as a writing aid. Any errors are our own.

\section*{Declarations}

Financial support was received from the Center for Social and Economic Behavior (C-SEB) at the University of Cologne and the German Research Foundation (DFG) under Germany’s Excellence Strategy, EXC 2126/1---390838866. In-kind support was also received from the U.S. National Science Foundation through TESS; data collected by Time-sharing Experiments for the Social Sciences, NSF Grant 0818839, Jeremy Freese and James Druckman, Principal Investigators. The author has no competing interests to declare that are relevant to the content of this article.
IRB approval was granted by the WiSo Ethics Review Board at the University of Cologne and the NORC. This project has been granted an exemption from ethics approval by the University of Melbourne (Project ID 36393).

\section*{Data availability}

All materials and data used in the analysis are freely available at \url{https://gitlab.com/gr0ssmann/undo}.

The experiment was preregistered at the Open Science Framework. See here for the preregistration: \url{https://osf.io/m7526}.

% \tableofcontents

\bigskip
\begin{center}
    $\diamond$
\end{center}
\bigskip

% !TeX spellcheck = en_US
\begin{quotation}
    Leibniz never married; he had considered it at the age of fifty; but the person he had in mind asked for time to reflect. This gave Leibniz time to reflect, too, and so he never married.
    \begin{flushright}
    ---\textit{Bernard de Fontenelle}
    \end{flushright}
\end{quotation}

\section{Introduction}
\label{secintro}

Formal rules such as man-made, positive laws have become ubiquitous, yet the relationship between different types of rules---soft and hard---is poorly understood.
This paper shows that soft and hard paternalistic rules are complements: waiting periods do not substitute for hard restrictions on behavior but are deployed as add-on restrictions.
We use a minimal, precisely preregistered survey experiment to estimate the relationship between \textit{(i)} limits to risky behavior and \textit{(ii)} mandatory waiting periods. One of these rules is softly paternalistic, while the other is an expression of hard paternalism. We contrast these rules to empirically investigate the political economy of the link between policy tools.
Moreover, we study the mental models underlying rule-making: how do beliefs about others' behavior shift when more time to decide is granted? If deliberation already changes behavior, does this reduce the need for strict paternalistic rules?

A large literature has studied the role of individual deliberation---here understood as time for reflection, not group discussion---on human decision-making. The idea that deliberative choices are different from, and perhaps superior to, affective ones, has reverberated through the ages. Behavioral economists have long conducted experiments and modeled behavior with exogenous variation in the time available for deciding between courses of action \eg{kocher2006time,caplin2016dual,caplin2016measuring,imas2022waiting}. \cite{loewenstein2015modeling} describe models (or ``systems'') of human behavior that are distinguished by the immediacy of choice. A related literature has studied decision-making under ``hot'' and ``cool'' states. \cite[3]{metcalfe1999hot} explicitly describe the ``cool system'' as ``contemplative'' and the ``hot system'' as ``impulsive.''

Governments have also purposely altered the temporal structure of decision-making, often with the stated intention of meliorating choices. Mandatory waiting periods are one such policy, found across many areas. While the U.S. federal government does not currently impose a waiting period for firearm purchases, several states do so \eg{luca2017handgun,edwards2018looking}. Jurisdictions in the U.S. and around the world require women seeking abortions to wait \eg{joyce2009impact,lindo2021new}. Sterilization \citep{rowlands2020mandatory}, marriage, divorce \citep{lee2013impact} and adoption are other areas where governments have imposed mandatory waiting periods. All of these laws decouple choice from action: spontaneous behavior is curtailed by giving decision-makers additional time to back out of a decision.
Similarly, the term ``cooling-off periods'' is used for laws that allow consumers to undo large purchases or contracts \citep{sher1967cooling,rekaiti2000cooling,sovern2013written}. The term is also used in negotiation protocols to avoid or resolve conflicts. \cite{oechssler2015cooling} use an online experiment to study the role of cooling-off periods in negotiations. Although these approaches differ from ours, they reflect a shared recognition that spontaneous behavior may differ from more contemplative behavior, sometimes to the detriment of decision-making quality.\footnote{\cite{fudenberg2018speed} modeled the joint distribution of choices and decision times, explaining empirical findings that agents that decide more quickly are more likely to choose correctly. However, in their model, decision times and choice quality are endogenous.}

We build on this idea for two purposes. First, we follow a recent line of investigation into behavioral determinants of the \emph{supply} of paternalism \citep{ambuehl2021motivates}.
In these studies, a decision-maker (``Chooser'') is matched with a policymaker (``Choice Architect'', CA). The CA can intervene in a decision faced by the Chooser.
Some of this work has modified the choice environment of Choosers to test how situational factors affect the degree of autonomy granted to them. For example, \cite{knf} varies the knowledge the Chooser has about a choice between two options. The more knowledge the Chooser possesses, the fewer interventions take place.

We modify how much time people have to think about a decision, and whether they can revise it.
We have Choosers participate in a simple decision that trades off risks against rewards: how many boxes to open in an adapted Bomb Risk Elicitation Task (BRET), an experimental game first proposed by \cite{crosetto2013bomb}.
In our experiment, Choosers can earn money by opening boxes shown on their screens. Each box earns the Chooser \$20, but one randomly selected box contains a ``curveball'' that eradicates all earnings.
This study uses the most extraordinary stakes ever reported in the extensive literature on the BRET.

We vary the temporal structure of decision-making and let CAs intervene by setting a cap on the number of boxes that can be opened by the Chooser, effectively constraining Choosers' risk-taking.
This allows us to test whether Chooser deliberation increases CA tolerance and what CAs believe about Choosers who have more time to think. The approach sheds light on the inherently subjective nature of rule-making and the mental models behind it.
That said, we address only one determinant of rule-making. Paternalism itself can be understood in many ways---for example, as a social preference anchored in preventing others from making losses, which are subjectively perceived by rule-makers \citep[e.g., employers:][]{buchmannpaternalistic}. Other motives for intervention remain possible, including motives that would not usually be considered paternalistic.

Second, we recognize that waiting periods and caps are two rules that, in practice, may be deployed simultaneously. For example, political scientists and environmental economists have studied ``policy mixes'' \eg{bouma2019policy}, a term also used by macroeconomists for the joint application of monetary and fiscal policy. The idea is that multiple instruments in a single policy area can attain policymakers' goals more effectively. Our experiment contains one treatment (EndoDelay) to test for this relationship. In this treatment, CAs can implement both the cap \emph{and} a waiting period. This tests for substitution between rules: do those CAs who implement a waiting period relax the cap? If deliberation is thought to move behavior in a preferred direction, hard-nosed restrictions may become less attractive if CAs generally respect autonomous decision-making.

Whether soft paternalism---such as waiting periods, which preserve choice---and hard paternalism---such as caps, which restrict it---substitute for or complement each other has direct implications for regulatory design (we discuss further nuances in Section~\ref{secsofthard}). \cite{tor2022law} notes that behavioral interventions can be combined with and perhaps supplant classic regulation. We provide an experimental test of this relationship. This paper contributes to experimental political economy by highlighting the inherently subjective nature of rule-making.
Our work also relates to the revision of choices. In economics \eg{NBERw28007,nielsen2022choices}, studies in this area have focused on disentangling intended but possibly nonstandard behavior from unintended mistakes. In social psychology \eg{gilbert2002decisions}, research has examined how behavior and attitudes are shaped by allowing people to change their minds. We address a different angle: how others respond to a decision-maker's ability to revise a choice.

In our experiment, decision-makers (Choosers) make a high-stakes risky decision about how many boxes to open in a task where opening too many risks losing all earnings. Other participants (Choice Architects, CAs) can set rules constraining Choosers' risk-taking.
First, we establish that most CAs believe that average Choosers open too many boxes relative to CAs' injunctive norm. Moreover, CAs tend to set the cap above their injunctive norm, replicating \citeauthor{pids}' finding (\citeyear{pids}) that CAs leave room for Choosers to express themselves.
Second, our highly powered survey experiment reveals that exogenous Chooser deliberation has no effect on the cap. Equivalence testing and Bayesian inference provide strong positive evidence for the null, ruling out even small effects.
This result is confirmed when considering CAs who were enabled to use both the cap and the waiting period: the exogenous availability of the waiting period does not cause a change in the cap, nor is the endogenous decision to implement it associated with a different cap.
Third, preregistered analyses reveal that deliberation makes CAs more optimistic about Chooser actions. This optimism is driven by CAs forecasting that average Choosers will take less risk when given time to deliberate.
We propose that soft and hard instruments target different parts of the population: waiting periods act on moderate Choosers, while caps constrain extreme ones.

\section{Soft and hard paternalism}
\label{secsofthard}

Gerald Dworkin's frequently cited definition of paternalism relates to an ``interference with a person's liberty of action justified by reasons referring exclusively to the welfare, good, happiness, needs, interests or values of the person being coerced'' \citep[65]{dworkin1972paternalism}. However, even policies that do not involve coercion yet attempt to meliorate choices---such as nudges \citep{thaler2008nudge}---are virtually universally viewed as inherently paternalistic. Indeed, Thaler and Sunstein speak of ``libertarian paternalism'' \eg{thaler2003libertarian,sunstein2003libertarian}, even though nudges explicitly do not restrict choice sets. Rather, nudges are thought to guide behavior in autonomy-preserving ways \citep{thaler2008nudge}. Many authors \eg{schnellenbach2012nudges,kirchgassner2017soft} have made a distinction between ``soft'' and ``hard'' paternalism, originally introduced by \cite{feinberg1989harm}. While some authors draw the distinction by referring to the ``costs'' imposed on a decision-maker \eg{sunstein2014nudge} or the strength of intervention \citep[687]{rizzo2009little}, Feinberg originally saw voluntariness as the deciding factor \citep{hanna2018hard}. If a decision would be ``voluntarily'' made---e.g., by a competent adult---interference is an expression of hard paternalism. On the other hand, the prevention of ill-informed or incompetent choices is softly paternalistic \citep[see also][]{mabsout2022john}. Providing information to uninformed decision-makers, and even preventing uninformed decisions \eg{knf}, may be viewed as softly paternalistic.

Setting aside the conceptual difficulties with Feinberg's distinction \eg{hanna2018hard}, waiting periods impose a delay between decision and action, allowing decision-makers to reflect and potentially reverse course. They are intended to prevent impulsive, ill-considered decisions---a rationale that appeals to Feinberg's delineation. Moreover, waiting periods are choice-preserving: they allow decision-makers to take any action. To the extent that waiting is less costly than a ban, they fit Sunstein's (\citeyear{sunstein2014nudge}) definition, too: ``Soft paternalism is weaker and essentially libertarian, in the crucial sense that it preserves freedom of choice'' (p. 19).
That said, waiting periods impose costs on decision-makers \citep[note 7]{loewenstein2007economist} and---since they affect all who wish to take an action, irrespective of their pre-existing level of consideration---are not asymmetrically paternalistic \citep{camerer2003regulation}. Feinberg's voluntariness-based criterion alone therefore does not fully classify them as soft, because the delay applies equally to those whose choices are already well-considered.
Similarly, \cite[18]{scoccia2018concept} writes that ``[some] nudges, such as mandatory `cooling off' periods for major financial or medical care decisions, may actually enhance autonomy''.
Laws that constrain behavior, on the other hand, remove autonomy in a hard paternalistic manner \eg{rizzo2009little}. Informed decision-makers face the cost of not being able to act on their true preferences, or even punishment for violations---hallmarks of hard paternalism. \cite[ch. 2]{sunstein2014nudge} discusses some of these nuances.

\section{Experimental design}
\label{secdesign}

\begin{figure}[t]
    \begin{center}
        \includegraphics[width=0.55\textwidth]{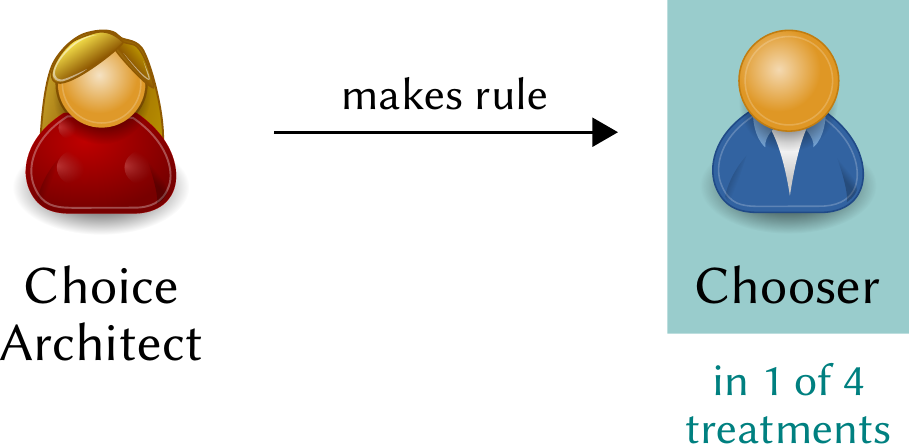}
    \end{center}

    \caption{Experimental design used in this paper}
    \label{cachooserdesignundo}
\end{figure}

Our design builds on the framework introduced by \cite{ambuehl2021motivates}. There are two types of participant: Choosers and Choice Architects (CAs). A Chooser is assigned to one of four treatments, and a CA is matched to him. The CA receives an explanation of the Chooser's experiment (including the treatment) and can then create a rule constraining the Chooser's actions (Figure \ref{cachooserdesignundo}).\footnote{In our experiment, CAs' decisions were stochastically implemented. There were four Choosers in total---one per treatment---and a CA was randomly selected for each of these Choosers to have their rule implemented. CAs knew that it was not certain whether their rule would be implemented, but all CAs had a strictly positive probability of making a rule for a real Chooser.} The experiment has a between design and is free of deception. We now describe the game faced by Choosers.

\subsection{Chooser experiment}
\label{choosergame}

Choosers participate in a two-day survey. At some point during these two days---as determined by treatment, see below---Choosers play the Bomb Risk Elicitation Task \citep[BRET,][]{crosetto2013bomb} with the highest stakes ever reported in the literature.
Our BRET works as follows: Choosers are faced with 25 boxes. Each box contains \$20 to be collected by the Chooser. Choosers decide how many (or, optionally, which specific) boxes to open; one randomly selected box contains a ``bomb.''\footnote{In our experiment, the word ``curveball'' was used instead of ``bomb,'' because the word ``bomb'' can carry negative associations. The Cambridge Dictionary lists the metaphorical use of the word ``curveball'' as implying, in American English, ``something such as a question or event that is surprising or unexpected, and therefore difficult to deal with'' (\url{https://dictionary.cambridge.org/dictionary/english/curveball}, accessed \today).}
If the box containing the ``bomb'' is among those opened, all earnings are eviscerated, leaving the Chooser with no payment from the BRET. Choosers learn these rules on day 1 of the survey.
While the BRET is ordinarily used to measure risk preferences, we use it to provide an intuitive game of risk taking.

\subsection{CA experiment}

\begin{figure}
    \centering

    \begin{tikzpicture}
        % Draw the horizontal line
        \draw[->] (0,0) -- (9,0);

        \fill[black] (1,0) circle (2pt);
        \node[anchor=south] at (1, 0.1) {\textbf{Start}};
        \node[anchor=north, align=center] at (1,-0.1) {Once$^{\text{I}}$\\Delay$^{\text{I}}$\\Undo$^{\text{I}}$\\EndoDelay$^{\text{I}}$};

        \fill[black] (4.5,0) circle (2pt);
        \node[anchor=north, align=center] at (4.5,-0.1) {Once$^{\text{D}}$\\~\\Undo$^{\text{D}}$\\EndoDelay$^{\text{D}\dagger}$};

        \fill[black] (8,0) circle (2pt);
        \node[anchor=south] at (8, 0.1) {\textbf{+1d}};
        \node[anchor=north, align=center] at (8,-0.1) {\phantom{Onothing$^{\text{X}}$}\\Delay$^{\text{D}}$\\Undo$^{\text{?}}$\\EndoDelay$^{\text{D}\dagger}$};
    \end{tikzpicture}

    \caption{Treatments}

    \caption*{\footnotesize $^{\text{I}}$Chooser obtains BRET instructions. $^{\text{D}}$Chooser decides. $^{?}$Optional revision to decision. $^{\dagger}$CA decides when decision is made. \emph{Note:} In all treatments, the Chooser experiment took two days.}

    \label{ufig1}
\end{figure}

We use a simple population-based survey experiment to test the hypotheses outlined below in Section \ref{sechypo}.\footnote{In this paper, we focus on the decisions and the experiment faced by CAs.} In such a survey, respondent samples representative of a target population can be used to combine the internal validity of experiments with the generalizability and lack of self-selection of a diverse pool of respondents \citep{mutz2011population}.
Our experiment contains four treatments, exogenously assigned to Choosers and hence to CAs (see Figure \ref{ufig1}).\footnote{Since each CA decided for only one Chooser in one particular treatment, it is accurate to say that CAs were ``assigned'' to the treatment that their Chooser was assigned to. However, recall that only some CAs' decisions were implemented.} Our treatments vary the temporal structure of the Chooser's decision-making.
In Once, the Chooser decides immediately after learning the rules of the game (Section \ref{choosergame}). In Delay, the Chooser has to wait one day between learning the game and actually opening the boxes. In Undo, the Chooser states a decision immediately after learning about the BRET, but can revise it on day 2; the chosen boxes are only opened on day 2. In EndoDelay, the CA decides whether the Chooser will be in Once or Delay. Since the Chooser experiment always takes two days, it is not possible for CAs to save Choosers time.

In \emph{all} treatments, CAs may set a cap on the number of boxes that can be opened by the Chooser---a hard paternalistic rule.
In EndoDelay, CAs additionally choose when the Chooser is to decide (on day 1 or day 2)---a softly paternalistic policy. This instrument implements deliberation as a pre-decision delay, unlike the post-decision revision opportunity seen in some real-world waiting periods (Section~\ref{secintro}); the Undo treatment captures that second channel. Thus, the first three treatments provide CAs with only the cap, while EndoDelay grants access to both instruments.

The CA experiment consisted of two survey pages.\footnote{English-language instructions are available in Section \ref{undoinstruct} of the Appendix.} The first page explained the Chooser's decision with a simple visualization and elicited an injunctive norm, beliefs about the behavior of the \emph{average} Chooser, and beliefs about Chooser happiness.
CAs received standard NORC AmeriSpeak survey compensation (a flat participation fee determined by NORC); beliefs about Chooser behavior were not separately incentivized.
Page 2 provided a treatment reminder and allowed CAs to make rules. CAs were told that to make no restrictions on the Chooser's behavior, they would have to enter ``25,'' the maximum number of boxes a Chooser can open. Beyond the treatment reminder, page 2 differed only for CAs in EndoDelay, who could also choose when the Chooser would decide. In all treatments except EndoDelay, four variables were elicited from CAs (injunctive norm, belief, happiness, cap). In EndoDelay, five were elicited (injunctive norm, belief, happiness, cap, when Chooser is to decide).

All instructions were optimized for general-population usage. The experiment was conducted in November 2023 in the English and Spanish languages using NORC AmeriSpeak through Time-sharing Experiments for the Social Sciences (TESS, \texttt{tessexperiments.org}). NORC AmeriSpeak is a well-established survey pool that provides representative, high-quality samples of the U.S. population. Unlike other survey pools, AmeriSpeak respondents are typically limited to participating in only a few surveys each month.
TESS enables researchers to access NORC AmeriSpeak for free through a competitive grant scheme.
IRB approval was granted by the WiSo Ethics Review Board at the University of Cologne and NORC, and the experiment was fully preregistered.
The Chooser experiment implemented the decisions of four CAs (one per treatment). It was conducted online in October 2024 at the Cologne Laboratory for Economic Research. This experimental protocol included box rewards stated in U.S. dollars, exactly as indicated to CAs (Section \ref{choosergame}). Official exchange rates by the European Central Bank were used to convert rewards into euros. The Chooser experiment was programmed in oTree \citep{chen2016otree} and spanned two days, with the identity of the box containing the curveball determined by a publicly verifiable lottery draw (Section~\ref{chooserprotocol}).

\section{Hypotheses}
\label{sechypo}

We use the final project proposal \citep[available in][a web resource]{undoproposal} to structure our research questions and hypotheses. The substance (though not the language) of the following hypotheses was submitted to TESS and approved through a refereeing process before the experiment was conducted. The proposal \citep[appendix C]{undoproposal} and the preregistration spell out in detail how to test these hypotheses econometrically. This constrains the exploratory nature of the study and makes inference more robust.

\subsection{Endogenously imposed waiting periods}
\label{sechypo1}

In EndoDelay, CAs have access to both the cap and the waiting period. The contrast with Once---where only the cap is available---allows us to test whether the availability of a soft intervention substitutes for hard imposition \citep{tor2022law}.
Several previous experiments on paternalism \eg{knf,pids,ambuehl2021motivates} have shown that CAs do not simply impose their own preference; they leave space for Choosers to express themselves \citep{pids}.
If a CA can choose between two means of achieving the same ends, she may well choose the less intrusive one. We do not claim that soft and hard measures are equally effective. Nonetheless, some CAs may shift to a less interventionist cap when they can simultaneously employ a less coercive tool.

\begin{hypothesis}\label{undohyp1}
    The availability of a soft intervention substitutes for hard intervention.
\end{hypothesis}

Moreover, we can compare the proportion of CAs who intervene in \emph{any} sense (hard, soft, or both) in EndoDelay to the proportion who intervene (necessarily hard) in Once. If the proportions are identical, the soft intervention could be viewed as a perfect substitute.

\begin{hypothesis}\label{undohyp2}
    The proportion of CAs who intervene in \emph{any} sense is independent of whether the soft intervention was available (that is, the soft intervention is a perfect substitute for hard intervention).
\end{hypothesis}

\subsection{Autonomy and deliberation}
\label{sechypo2}

Treatments Once, Delay and Undo allow us to investigate how CAs' rule-making responds to \emph{exogenous} variation in Choosers' deliberation time, holding the available instrument (the cap) fixed. We study three cases: \textit{(i)} Choosers have no additional time (Once); \textit{(ii)} Choosers have one day to think about their decision (Delay); \textit{(iii)} Choosers decide immediately but can optionally revise their decision one day later (Undo).
First, we test whether deliberation causes CAs to ``go easy'' on the cap. In other words, does contemplation cause more respect for autonomy?

\begin{hypothesis}\label{undohyp3}
    The mean cap set in case \textit{(ii)} is higher than in case \textit{(i)}.
\end{hypothesis}

Second, researchers have long recognized status-quo bias in human decision-making \citep{samuelson1988status}. In case \textit{(iii)}, Choosers may simply forget to correct an unfavorable initial choice. That alone would make it closer to case \textit{(i)}. Hence, we test whether CAs recognize this possibility.

\begin{hypothesis}\label{undohyp4}
    The mean cap set in case \textit{(ii)} is higher than in case \textit{(iii)}.
\end{hypothesis}

Third, while case \textit{(iii)} is perhaps inferior to pure deliberation, it does allow for correction.

\begin{hypothesis}\label{undohyp5}
    The mean cap set in case \textit{(iii)} is higher than in case \textit{(i)}.
\end{hypothesis}

In sum, hypotheses \ref{undohyp3}--\ref{undohyp5} predict a ranking of (mean) caps: the most restrictive caps are expected for case \textit{(i)} and the least restrictive for case \textit{(ii)}, with case \textit{(iii)} in between the extremes.
Our experimental treatments Once, Undo and Delay enable direct tests of these hypotheses.

\subsection{Mental models}

We now turn to the beliefs that underlie CAs' rule-making. We focus on the cleanest contrast---Once \textit{vs.} Delay, i.e., cases \textit{(i)} and \textit{(ii)}---because Undo involves both immediate decision-making and a revision opportunity, making it a hybrid of the two temporal structures. First, does deliberation cause CAs to believe that Choosers come closer to CAs' subjectively preferred bliss point?

\begin{hypothesis}\label{undohyp6}
    The error\footnote{The error is defined below (equation \ref{undoerror}) as the absolute difference between a CA's injunctive norm and her beliefs about average Chooser behavior.} is lower in case \textit{(ii)} than in case \textit{(i)}.
\end{hypothesis}

Second, we elicit CAs' beliefs about Choosers' happiness.

\begin{hypothesis}\label{undohyp7}
    The mean forecast happiness (mean expected norm deviation) is higher in case \textit{(ii)} than in case \textit{(i)}.
\end{hypothesis}

These mechanisms allow us to investigate how CAs' mental models depend on Chooser deliberation.

\section{Results}
\label{secresults}

We first discuss descriptive statistics from the CA experiment, then present results from preregistered analyses (detailed in the Appendix) alongside additional Ordinary Least Squares regression models. The regression models do not contradict the preregistered analyses. We also provide robustness checks and alternative specifications. All regression models use HC3 heteroskedasticity-consistent standard errors.

CAs could skip individual items, but the survey asked respondents to reconsider before submitting an empty response.
We include CAs for whom data on the cap are available or, in EndoDelay, data on both the cap and the Chooser's day to decide. From an original sample of 2,714 CAs, this preregistered filtering procedure leaves 2,702 (Once: 639, EndoDelay: 690, Delay: 717, Undo: 656). Attrition is thus not a significant concern and is balanced across treatments. Whenever we report on injunctive norms, beliefs, or happiness, the sample size is slightly reduced: in 39 cases (1.44\%), the injunctive norm is missing; in 41 cases (1.52\%), the belief is missing; in one case, happiness is missing.

\subsection{Caps, norms and beliefs}
\label{undodescr}

Across all treatments, the mean cap is 17.5 boxes (std.dev.: 7.6). The mean (injunctive) norm is 9.6 boxes (std.dev.: 5.9). The mean belief about the number of boxes opened by the \emph{average} Chooser is 11.4 (std.dev.: 5.7).
38.7\% of CAs in EndoDelay imposed the 1-day waiting period.

In 92.8\% of cases where the norm is available, the cap is weakly larger than the norm; in 74.8\%, the cap is strictly larger.
Moreover, in 77.2\% of cases where both norm and belief are available, CAs believe that the average Chooser opens weakly more boxes than the CA perceives he should open. In 57.2\% of cases, the inequality is strict.
These patterns confirm that an upper limit was the right design choice.

\begin{result}\label{undoopentoomany}
    CAs believe that the average Chooser opens too many boxes (relative to CAs' perceived optimal choice).
\end{result}

\begin{figure}
    \includegraphics[width=\textwidth]{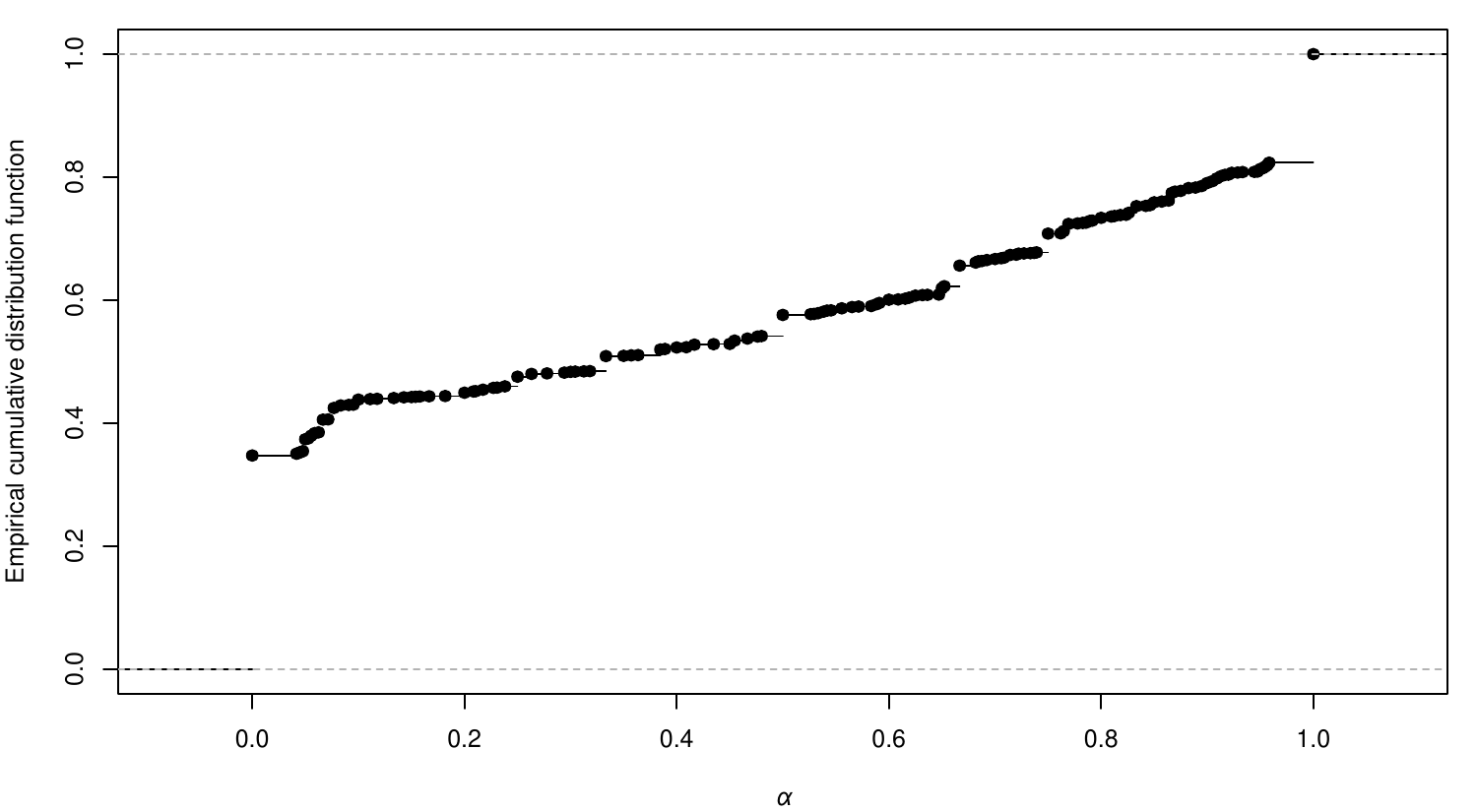}

    \caption{Distribution of $\alpha$}
    \caption*{\footnotesize This Figure is restricted to cases where $\alpha$ is defined and in $[0, 1]$.}
    \label{alphaundo}
\end{figure}

For each CA $i$, we can define $\alpha_i$ as the solution to the following equation:

\begin{equation*}
    \text{Cap}_i = \alpha_i \text{Norm}_i + (1-\alpha_i) \cdot 25.
\end{equation*}

The distribution of $\alpha$ across CAs captures how they balance their injunctive norm against imposing no restriction.
In 89.6\% of cases, $\alpha$ is defined and in $[0, 1]$. Figure \ref{alphaundo} shows the distribution of $\alpha$ in this subset. $\alpha = 0$ represents the modal choice (reflecting no cap), attained in about one third of cases, and $\alpha = 1$ is the second most common $\alpha$. The median $\alpha$ is 0.333 and the mean is 0.42. In sum, while about two thirds of CAs intervene, Choosers are generally put on a long leash---a pattern previously documented in the literature.

\begin{result}
    As in \cite{pids}, CAs tend to leave room for Choosers to express themselves.
\end{result}

For the analysis below, we define one additional preregistered variable that relates to beliefs about Chooser behavior and the injunctive norm:

\begin{equation}\label{undoerror}
    \text{Error}_i = \left| \text{Belief}_i  - \text{Norm}_i \right|.
\end{equation}

This variable captures the absolute expected norm deviation of the \emph{average} Chooser. It is bounded from below by 0. The mean error is 4.6 (std.dev.: 4.2). The notion of an ``error'' here entirely reflects the subjective judgment of CAs.
In the Appendix, we also use $1\left[\text{Error}_i = 0\right]$ as an outcome variable.

\subsection{Hard paternalism and deliberation}

\begin{table}
    \begin{center}
        \sisetup{parse-numbers=false, table-text-alignment=center}
        \begin{adjustbox}{max width=\textwidth}%
        \begin{threeparttable}
                \begin{tabular}{l S[table-format=4.6] S[table-format=4.6] S[table-format=4.6] S[table-format=4.6] S[table-format=3.6] S[table-format=4.6]}
                    \toprule
                    & {Model 1} & {Model 2} & {Model 3} & {Model 4} & {Model 5} & {Model 6} \\
                    \midrule
                    Intercept       & 17.358^{***} & 9.590^{***} & 12.158^{***} & 5.005^{***} & 17.638^{***} & 0.321^{***} \\
                    & (0.294)      & (0.232)     & (0.225)      & (0.166)     & (0.364)      & (0.018)     \\
                    Undo            & -0.067       & -0.283      & -1.291^{***} & -0.743^{**} &              & -0.017      \\
                    & (0.423)      & (0.327)     & (0.325)      & (0.240)     &              & (0.026)     \\
                    Delay           & 0.329        & 0.252       & -1.024^{***} & -0.607^{**} &              & 0.032       \\
                    & (0.409)      & (0.323)     & (0.306)      & (0.230)     &              & (0.026)     \\
                    EndoDelay       & 0.333        & 0.016       & -0.531       & -0.290      &              & 0.034       \\
                    & (0.412)      & (0.324)     & (0.315)      & (0.229)     &              & (0.026)     \\
                    Waiting imposed &              &             &              &             & 0.137        &             \\
                    &              &             &              &             & (0.598)      &             \\
                    \midrule
                    Outcome         & {Cap}        & {Norm}      & {Belief}     & {Error}     & {Cap}        & {Cap is 25} \\
                    Subset          & {---}        & {---}       & {---}        & {---}       & {EndoDelay}  & {---}       \\
                    R$^2$           & 0.001        & 0.001       & 0.007        & 0.005       & 0.000        & 0.002       \\
                    Adj. R$^2$      & -0.001       & -0.000      & 0.006        & 0.003       & -0.001       & 0.001       \\
                    Num. obs.       &{2702}        &{2663}       &{2661}        &{2649}       &{690}         &{2702}       \\
                    \bottomrule
            \end{tabular}
            \begin{tablenotes}[flushleft]
                \scriptsize{\item OLS estimates. HC3 standard errors in parentheses. Reference category: Once. Model~5 restricted to EndoDelay. $^{***}p<0.001$; $^{**}p<0.01$; $^{*}p<0.05$.}
            \end{tablenotes}
        \end{threeparttable}
        \end{adjustbox}
        \caption{Outcomes by treatment}
        \label{tu1}
    \end{center}
\end{table}

\begin{figure}
    \includegraphics[width=\textwidth]{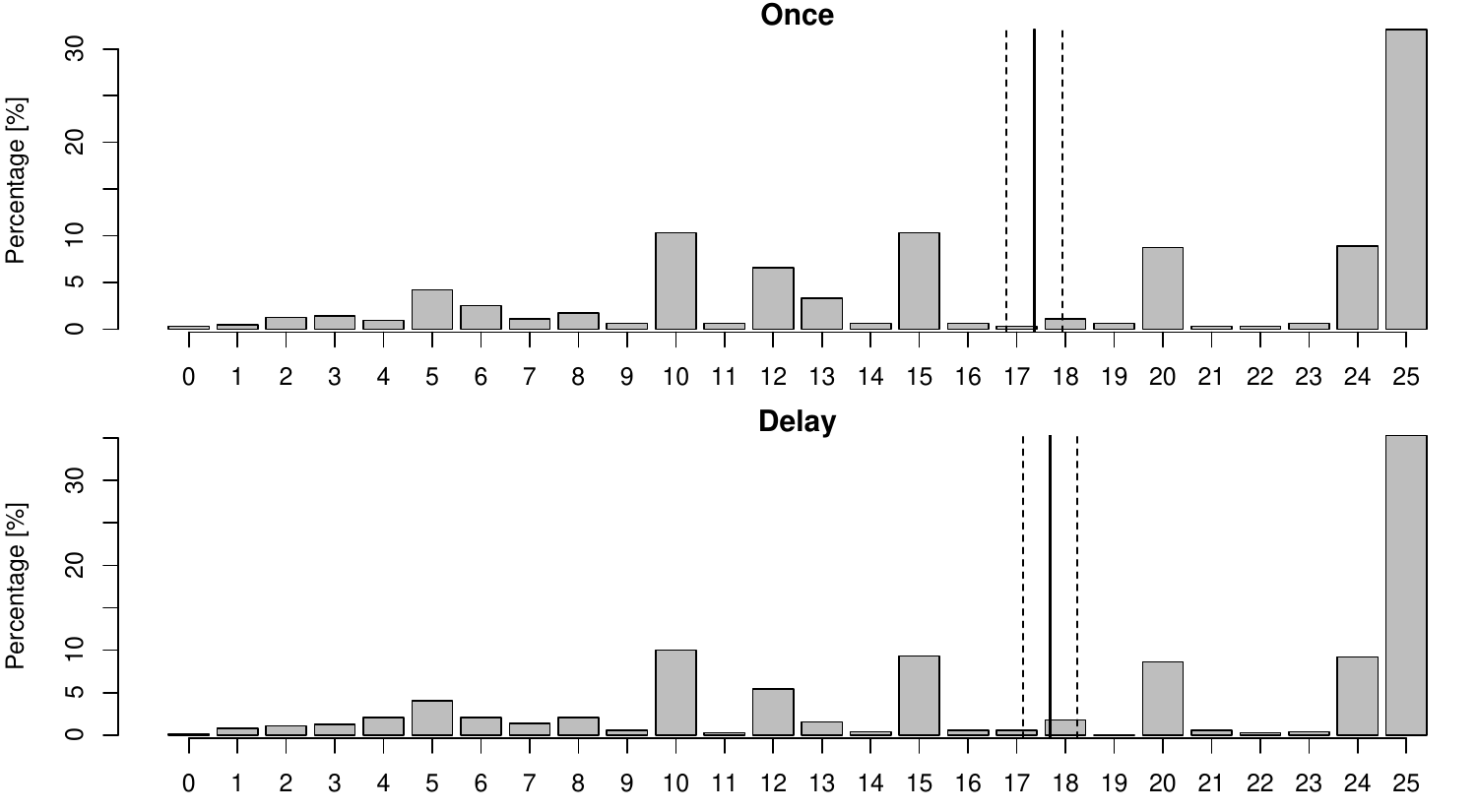}

    \caption{Distribution of caps in Once and Delay}
    \caption*{\footnotesize Vertical lines show sample means. 95\% confidence interval based on one-sample $t$-tests.}
    \label{undoplot2}
\end{figure}

Figure \ref{undoplot2} suggests that the distribution of caps does not differ significantly between two key treatments. Model 1 in Table \ref{tu1} confirms this for all treatments, as do the preregistered analyses in Sections \ref{undopra1}, \ref{undopra3}, \ref{undopra4} and \ref{undopra5} of the Appendix.

We can go beyond simply failing to reject: focusing on the cleanest comparison---Once \textit{vs.} Delay (hypothesis \ref{undohyp3})---we apply a TOST equivalence procedure \citep{lakens2018equivalence} and a Bayesian $t$-test \citep{rouder2009bayesian} to make a positive case for the absence of an effect.
The difference in mean caps is $-0.329$ ($\text{SE} = 0.409$). The corresponding 90\% confidence interval, $[-1.002, 0.343]$, translates to $[-0.133, 0.046]$ in Cohen's $d$ units (pooled $\text{SD} = 7.520$). Because this interval falls entirely inside $[-0.2, 0.2]$---the conventional smallest effect size of interest---we reject effects of even small magnitude at $\alpha = 0.05$ in the TOST framework. In substantive terms, $d = 0.2$ corresponds to roughly 1.5 boxes---about 6\% of the 25-box range---implying a shift in the cap sufficient to alter the Chooser's maximum expected earnings by \$30.
A complementary Bayesian independent-samples $t$-test with a Cauchy($0, r\!=\!\frac{\sqrt{2}}{2}$) prior on effect size returns $\text{BF}_{01} = 11.9$, meaning the data are roughly 12 times more probable under the null than under the alternative---strong evidence for the absence of a treatment effect according to conventional classification \citep{lee2013bayesian}. The result is insensitive to the prior width: narrower ($r = 0.5$: $\text{BF}_{01} = 8.5$) and wider ($r = 1.0$: $\text{BF}_{01} = 16.7$; $r = 2.0$: $\text{BF}_{01} = 33.4$) priors all favor the null.

A robustness check in model 6 shows that this null result also holds for the binary outcome of whether the CA sets no cap at all, addressing the concern that mean equivalence could mask changes at the focal value of 25. The proportion of CAs setting no cap does not change between Once, Undo and Delay (Section \ref{undopra10} in the Appendix).

\begin{result}\label{undoresult1}
    Chooser deliberation does not cause a change in the cap. Hypotheses \ref{undohyp1}, \ref{undohyp3}, \ref{undohyp4} and \ref{undohyp5} are rejected (as null results at high power).
\end{result}

We also elicited injunctive norms and beliefs. As expected, injunctive norms do not differ between treatments (model 2). We discuss beliefs and error (models 3--4) below.

\subsection{Waiting periods as add-on restrictions}

Model 5 in Table \ref{tu1} shows that CAs in EndoDelay who actually implemented the waiting period do not relax the cap. This reinforces result \ref{undoresult1}. Beyond that, this finding demonstrates how \emph{even the conscious selection of a 1-day delay does not substitute for hard imposition}.

Since we find no substitution between these two policies, waiting periods represent an ``add-on restriction'' in our setting: they are an additional rule placed onto Choosers. The preregistered analysis in Section \ref{undopra2} of the Appendix confirms this result as highly statistically significant ($p < 0.001$). Simply put, the only difference between Once and EndoDelay is that the latter also allows imposing the waiting period. 67.9\% of CAs in Once imposed a cap of strictly less than 25. Almost exactly the same proportion, 64.5\% of CAs in EndoDelay, imposed a cap of strictly less than 25. 38.7\% of CAs in EndoDelay imposed the waiting period, with 25.2\% implementing both the waiting period and a cap, while 13.5\% \emph{only} implemented the waiting period. 39.3\% \emph{only} implemented the cap. In sum, 78.0\% of CAs in EndoDelay intervened in \emph{any} way. Since the cap does not give way to the waiting period, the possibility of imposing the waiting period merely increases the number of CAs that intervene at all, and represents a policy dimension distinct from the cap. Since our design imposes no cost for combining instruments, this pattern demonstrates non-substitutability under a costless menu expansion.

\begin{result}
    Waiting periods are add-on restrictions. Hypothesis \ref{undohyp2} is rejected.
\end{result}

The preregistered analyses in Section \ref{undopra9} of the Appendix further suggest that CAs in EndoDelay, conditional on their choice between Once and Delay for the Chooser, behave similarly to CAs exogenously assigned to Once or Delay.

\subsection{Mental models of deliberation}
\label{undomentmodresult}

\begin{figure}
    \includegraphics[width=\textwidth]{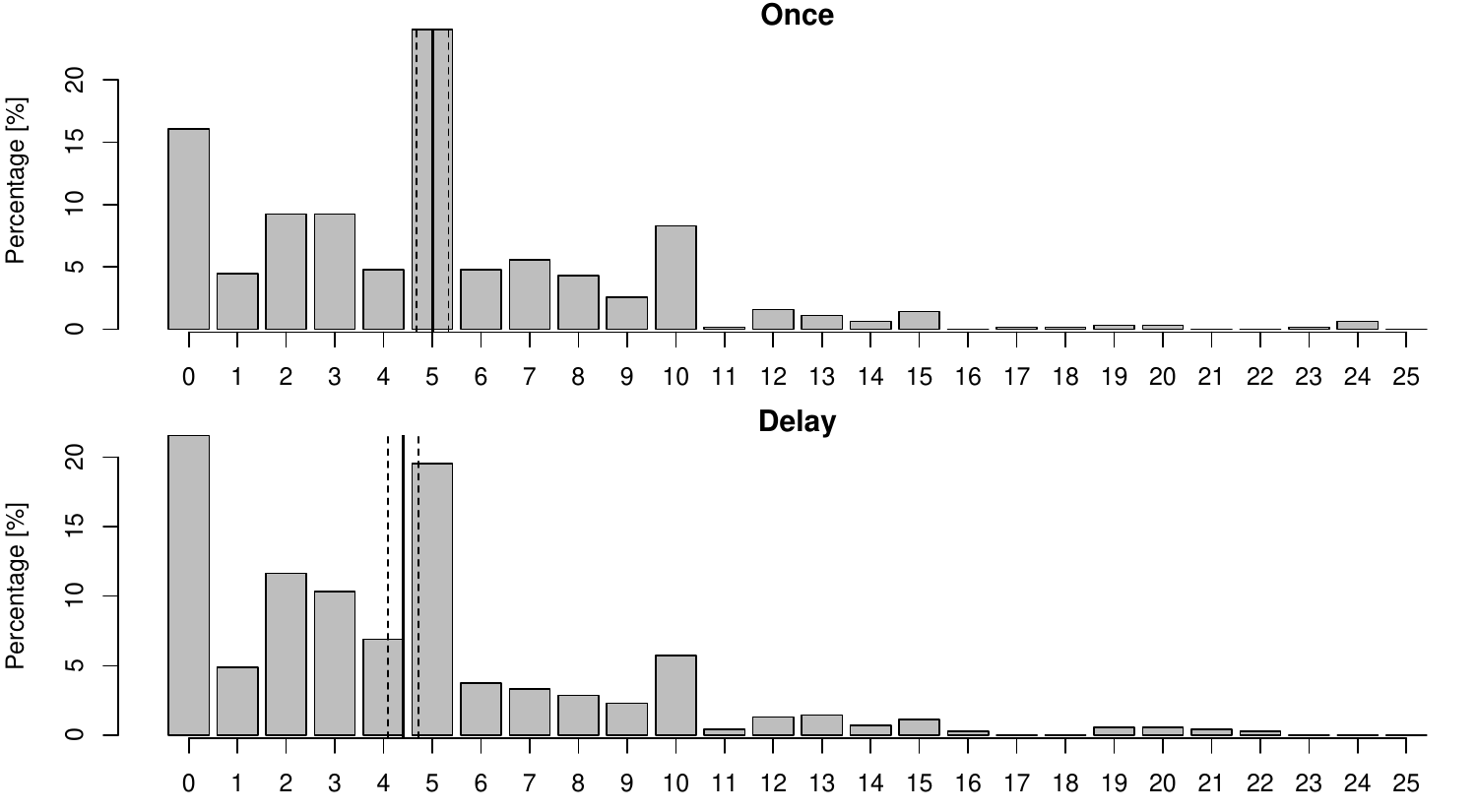}

    \caption{Distribution of expected errors in Once and Delay}
    \caption*{\footnotesize Vertical lines show sample means. 95\% confidence interval based on one-sample $t$-tests.}
    \label{undoplot3}
\end{figure}

As suggested by Figure \ref{undoplot3} and models 3--4 in Table \ref{tu1}, beliefs about average behavior and average errors (Equation \ref{undoerror}) are causally reduced by deliberation, confirmed by the preregistered analysis in Section \ref{undopra7} of the Appendix ($p = 0.009$; adjusted $p = 0.017$). Section \ref{undomht} in the Appendix discusses the adjustment for multiple hypothesis testing.
Model 1 in Table \ref{tu5} (Appendix) replicates this for the binary outcome of zero error.
Deliberation reduces errors.
As expected, the injunctive norm is unchanged across treatments (model 2 in Table \ref{tu1} and Section \ref{undopra6} in the Appendix).

\begin{result}
    CAs believe that Choosers, with deliberation, come closer to CAs' subjectively preferred choice. Hypothesis \ref{undohyp6} (mechanism 1) is confirmed.
\end{result}

We now turn to exploratory analyses distinguishing two mechanisms that could drive this reduction in error.
Recall that most CAs believe that the average Chooser opens too many boxes (result \ref{undoopentoomany}). Because error is defined as $|\text{Belief} - \text{Norm}|$ (Equation \ref{undoerror}), a reduction in error can arise through at least two distinct channels:

\paragraph{Story 1: Convergence toward the norm.} CAs believe that deliberation causes Choosers to move closer to the CA's own bliss point, regardless of where that bliss point lies. Under this mechanism, a CA whose norm is high (e.g., 20 boxes) would predict that deliberation causes the average Chooser to open \emph{more} boxes, while a CA whose norm is low (e.g., 5) would predict a decrease. The key testable implication is that the CA's norm \emph{moderates} the treatment effect on beliefs. Note that, in a level specification, the resulting error reduction is proportional to baseline error, which itself varies with the norm; norm moderation of the effect on errors can therefore also arise under this story.

\paragraph{Story 2: Uniform reduction in risk-taking.} CAs believe that deliberation simply causes fewer boxes to be opened, independent of the CA's norm. The aggregate reduction in error is a byproduct: since most CAs think that Choosers open too many boxes, a uniform downward shift in beliefs mechanically reduces the average error. However, for CAs whose norm is itself high, this downward shift can actually \emph{increase} the distance between belief and norm. Hence, this story also predicts norm moderation of the effect on errors, but---unlike story~1---does \emph{not} predict norm moderation of the treatment effect on beliefs.

\medskip\noindent Both stories can generate norm moderation of the treatment effect on errors in a level specification, so errors alone cannot distinguish them. The stories diverge on beliefs: only story~1 predicts that the CA's norm moderates the treatment effect on beliefs. This yields the key test.

\begin{table}
    \begin{center}
        \sisetup{parse-numbers=false, table-text-alignment=center}
        \begin{adjustbox}{max width=\textwidth}%
        \begin{threeparttable}
                \begin{tabular}{l S[table-format=4.6] S[table-format=4.6] S[table-format=4.6] S[table-format=4.6]}
                    \toprule
                    & {Model 1} & {Model 2} & {Model 3} & {Model 4} \\
                    \midrule
                    Intercept                   & 11.883^{***} & 7.512^{***} & 4.854^{***} & 4.167^{***}  \\
                    & (0.158)      & (0.291)     & (0.114)     & (0.269)      \\
                    Delay or Undo               & -0.877^{***} & -0.926^{*}  & -0.521^{**} & -1.309^{***} \\
                    & (0.221)      & (0.410)     & (0.164)     & (0.381)      \\
                    Norm                        &              & 0.456^{***} &             & 0.072^{**}   \\
                    &              & (0.030)     &             & (0.028)      \\
                    Delay or Undo $\times$ Norm &              & 0.006       &             & 0.082^{*}    \\
                    &              & (0.042)     &             & (0.039)      \\
                    \midrule
                    Outcome                     & {Belief}     & {Belief}    & {Error}     & {Error}      \\
                    R$^2$                       & 0.006        & 0.229       & 0.004       & 0.032        \\
                    Adj. R$^2$                  & 0.005        & 0.228       & 0.003       & 0.031        \\
                    Num. obs.                   &{2661}        &{2649}       &{2649}       &{2649}        \\
                    \bottomrule
            \end{tabular}
            \begin{tablenotes}[flushleft]
                \scriptsize{\item OLS estimates. HC3 standard errors in parentheses. ``Delay or Undo'' pools the Delay and Undo treatments; the reference group pools Once and EndoDelay. $^{***}p<0.001$; $^{**}p<0.01$; $^{*}p<0.05$.}
            \end{tablenotes}
        \end{threeparttable}
        \end{adjustbox}
        \caption{Heterogeneous treatment effects}
        \label{tu2}
    \end{center}
\end{table}

To adjudicate between the two stories, we fitted the models in Table \ref{tu2}. This table groups Delay and Undo together, and Once and EndoDelay together, to increase power for estimating heterogeneous treatment effects.\footnote{Recall that Once and EndoDelay had an identical page 1 in the survey---where the norm and beliefs were elicited---and Delay and Undo share the crucial feature of Chooser deliberation. Table \ref{tu4} in the Appendix gives results for all four treatments separately, confirming that Undo and Delay effects go in the same direction.}
In all models, the addition of CAs' injunctive norm adds explanatory power, a kind of false consensus effect \citep{ross1977false}.

Model 2 reveals \emph{no} interaction between deliberation and the norm on beliefs. On average, CAs forecast that deliberation causes a uniform decrease in the number of boxes opened, regardless of their own norm. This is inconsistent with story 1, which predicts that high-norm CAs would forecast an increase in boxes opened under deliberation.
Turning to errors, model 4 shows that the effect of deliberation \emph{is} moderated by the norm ($p = 0.038$). While both stories predict such moderation, the direction matches story~2. \textit{Ceteris paribus}, the higher a CA's injunctive norm, the smaller the error reduction through deliberation. Beyond a norm of about $16$, deliberation even causes a predicted \emph{increase} in error---as story 2 predicts, because a uniform downward shift in beliefs moves Choosers \emph{away from} a high-norm CA's bliss point. On average, CAs estimate that deliberation causes a reduction in the number of boxes opened by Choosers, and because most CAs believe that too many boxes would be opened without deliberation (Section \ref{undodescr}), this shift also reduces the error for the typical CA.
Model 2 in Table \ref{tu3} in the Appendix replicates this result for the binary outcome of the error being nil.

These results should be interpreted with caution, however, as they partly depend on grouping treatments. In Table \ref{tu4} in the Appendix, we report the ungrouped models. While the absence of norm moderation on beliefs (model 2) is confirmed---no treatment shows a significant interaction, supporting story 2---the evidence on errors is less clear-cut. Only the Delay $\times$ Norm interaction on errors is individually significant in model 4 of Table \ref{tu4} ($p < 0.05$); the corresponding Undo $\times$ Norm interaction is not, although the coefficient points in the same direction. Moreover, model 2 in Table \ref{tu5} in the Appendix shows that the interaction on the binary outcome (whether the error is nil) does not obtain without treatment grouping. These null results in the ungrouped specifications may reflect insufficient power to detect a small interaction in individual treatments rather than genuine absence of the pattern.
We therefore view the evidence as suggestive of story 2 for the average CA, while acknowledging that the discrimination between the two stories is not definitive in the ungrouped specifications.

\begin{result}
    On average, CAs believe that Choosers come closer to CAs' injunctive norm under deliberation because deliberation is believed to reduce the number of boxes opened and CAs tend to believe that the average Chooser opens too many boxes (result \ref{undoopentoomany}). This pattern supports a uniform shift in beliefs (story 2) rather than convergence toward CAs' individual norms (story 1), although the evidence depends in part on the grouping of treatments.
\end{result}

Section \ref{undopra8} of the Appendix shows that forecast Chooser happiness does not differ significantly between Once and Delay. We do not report these results in the main text.

\section{Discussion and conclusion}
\label{secconclusion}

Our results present a puzzling empirical picture. CAs do not react to Choosers' deliberation when setting limits to risk-seeking behavior, nor do they substitute between soft and hard paternalistic rules---yet they believe that Choosers' behavior is changed by deliberation. As shown in Section \ref{secresults}, this null effect is not merely a failure to reject: equivalence testing and a Bayesian analysis converge on strong evidence that the true effect is negligibly small.

Several design features limit how far these findings can be generalized. The BRET presents a cold, calculative risk environment (although we use the highest stakes reported in the literature). The null effect of deliberation on caps may not generalize to affective or visceral domains---such as firearm purchases or abortion decisions---where waiting periods are most commonly deployed and where additional deliberation time might more plausibly revise behavior. Moreover, our design imposes no administrative or welfare cost on the Choice Architect for deploying multiple instruments simultaneously, and any individual CA's probability of being the one implemented was small. In real-world policy settings, where combining rules entails budget constraints or accountability costs, the observed joint deployment could be attenuated.

Our design is silent on whether CAs who \emph{implemented} the waiting period are more optimistic about changes in behavior. However, we can use the results from treatments Once, Delay and Undo to form a prior expectation. Deliberation causes a shift in CAs' beliefs about the behavior of the \emph{average} Chooser, yet hard interventions remain unchanged. Most CAs believe that Choosers open too many boxes; and caps on Choosers' behavior are generally above beliefs about average behavior. These findings suggest that soft and hard interventions---waiting periods and limits, in our case---target different parts of the population. Soft interventions target Choosers that are misguided but within a close-to-normal range of behavior. Hard interventions target ``extreme'' Choosers. This explains the simultaneous use of waiting periods and the cap and their non-substitutability. Because this account rests in part on the exploratory, treatment-grouped analysis in Section~\ref{undomentmodresult}, we cannot confirm it with the present data. Moreover, beliefs were elicited only about the \emph{average} Chooser, leaving the upper tail of the distribution---where hard interventions are hypothesized to bind---unmeasured. In EndoDelay, beliefs were elicited on page~1, before CAs decided on page~2 whether to impose the waiting period, so we lack a direct measure of what delay-adopters believe a delayed Chooser would do.

To our knowledge, this point has not been explicitly recognized in the paternalism literature, although heterogeneous effects with respect to the targeted outcome itself have been described. One example concerns the effect of alcohol taxes on drinkers' consumption depending on where they lie in the distribution of drinkers \citep{manning1995demand}.
Similarly, researchers have recently shown how heterogeneous decision-makers can be targeted if the effects of some policy on their individual behavior can be estimated \citep{lipman2024one,opitz2024algorithmic,lipman2024tailoring}.
The behavioral theory behind policy targeting here is simple: waiting may be thought to avert smaller mistakes, but tougher restrictions are required to prevent larger norm deviations.

Recent work by \cite{allcott2022nudges} has emphasized that non-standard policies like nudges can enhance behavioral distortions. Given its limited scope, our experiment does not reveal much about the motives of CAs in EndoDelay who chose to implement the waiting period (or not). CAs could not save Choosers time---as the survey always took exactly two days---and yet, a majority of CAs did not choose to implement the waiting period. Were they pessimistic about waiting periods' effects? What downsides do waiting periods have? Our experiment cannot answer these questions.

The non-invasive, choice-preserving nature of waiting periods has long been used as an argument in their favor.
In \textit{Planned Parenthood v. Casey} (\citeyear{casey1992}), the U.S. Supreme Court established that no ``undue burden'' could be placed on abortion access, and waiting periods generally passed this test.
\cite[380]{kalmanson2016second} discusses the ``undue burden'' jurisprudence with respect to firearms regulation and restrictions on abortion.
However, now that \textit{Roe v. Wade} (\citeyear{roe1973}) and \textit{Casey} have been overturned, states can ban abortion outright, and hard restrictions may substitute for softer waiting periods.

While our results offer some insight into CAs' mental models, we can only speculate about their motivations in choosing between policy instruments. If softly paternalistic measures do not substitute for hard paternalism, innovations in policy instruments may increase the total regulatory burden. Policymakers can simultaneously apply diverse rules to target specific parts of society, even within the same policy area.

\bibliography{bib}

\appendix

\parindent0mm
\parskip1em

\section{Survey items}
\label{undoinstruct}

\subsection{Page 1}

\begin{center}
    \includegraphics[width=\textwidth]{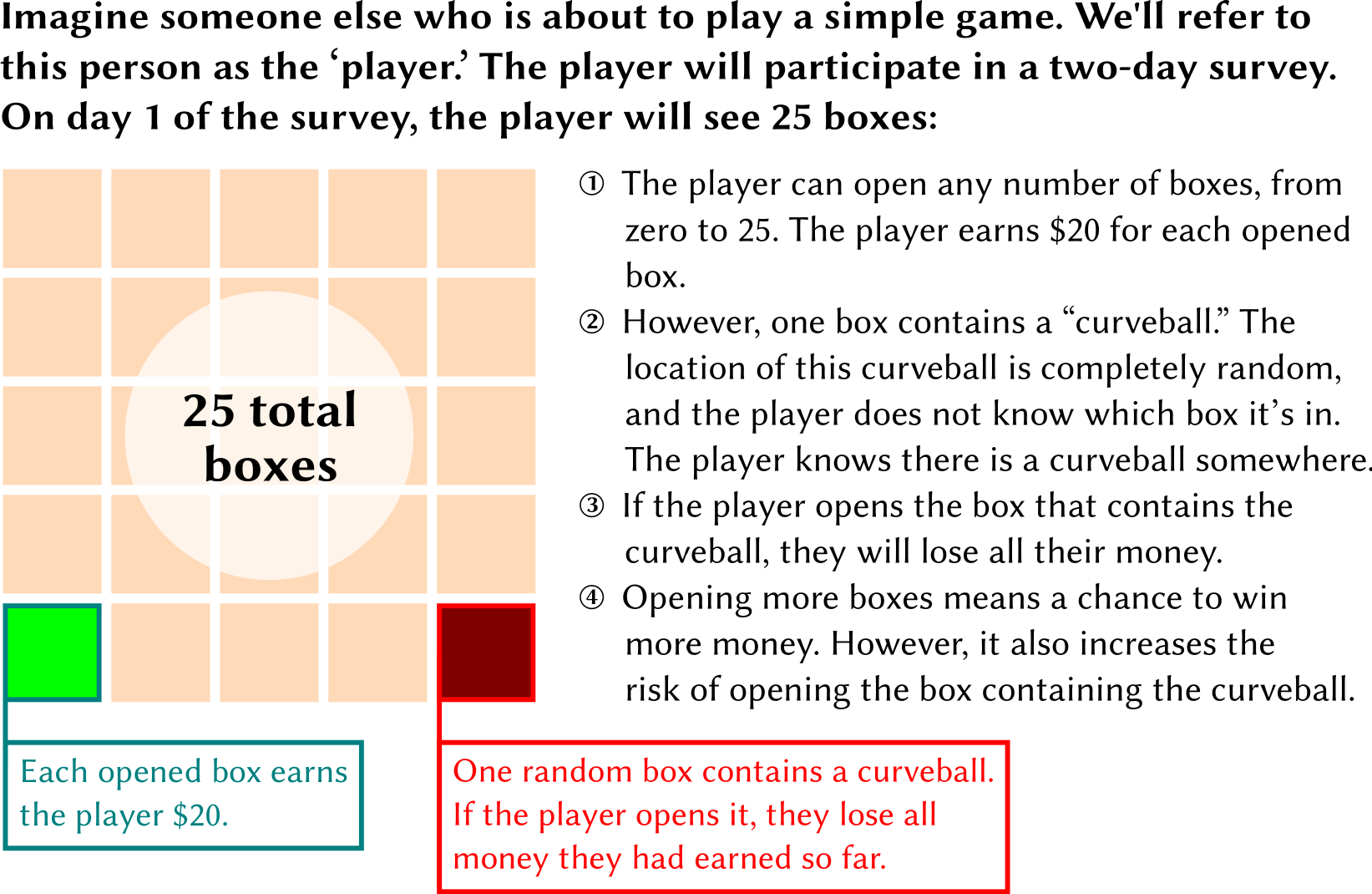}
\end{center}

\begin{center}
    \begin{tabular}{@{}p{0.225\textwidth}p{0.225\textwidth}p{0.225\textwidth}p{0.225\textwidth}@{}}
        \toprule
        \multicolumn{1}{c}{\textbf{Once}} & \multicolumn{1}{c}{\textbf{EndoDelay}} & \multicolumn{1}{c}{\textbf{Delay}} & \multicolumn{1}{c}{\textbf{Undo}} \\ \midrule
        On day 1, the player will learn the rules of the game and decide right away which boxes to open. & \emph{(as in Once)} & On day 1, the player will learn the rules of the game and decide one day later (on day 2) which boxes to open. & On day 1, the player will learn the rules of the game and decide right away which boxes to open. One day later (on day 2), they are allowed to change their decision or stick with their initial decision. The boxes that are chosen will be opened on day 2. \\ \bottomrule
    \end{tabular}
\end{center}

In your personal opinion, how many boxes should the player open?

[Numeric input field, allowable values: 0, 1, …, 25]

Remember: The survey always takes two days. Assume that the player participates on both days. There is no way to shorten the survey.

What do you think: How many boxes will the average player ultimately open? Tell us your best guess.

[Numeric input field, allowable values: 0, 1, …, 25]

How happy do you think the average player will feel after making their choice? Rate the player's expected happiness on a scale:

[Numeric input field, Likert, radio buttons, allowable values: 1 (extremely unhappy) to 7 (extremely happy)]

\par\vspace{-.5\ht\strutbox}\noindent\hrulefill\par

\begin{center}
    \textbf{Pagebreak}
\end{center}

\par\vspace{-.5\ht\strutbox}\noindent\hrulefill\par

\subsection{Page 2}

Today, you are granted the power to either let the player have their choice or set a maximum number of boxes they can open.

\begin{center}
    \begin{tabular}{@{}p{0.225\textwidth}p{0.225\textwidth}p{0.225\textwidth}p{0.225\textwidth}@{}}
        \toprule
        \multicolumn{1}{c}{\textbf{Once}} & \multicolumn{1}{c}{\textbf{EndoDelay}} & \multicolumn{1}{c}{\textbf{Delay}} & \multicolumn{1}{c}{\textbf{Undo}} \\ \midrule
        Remember, on day 1, the player will learn the rules of the game and decide right away which boxes to open. & You will also decide when they are to make the choice. Remember, on day 1, the player will learn the rules of the game. & Remember, on day 1, the player will learn the rules of the game and decide one day later (on day 2) which boxes to open. & Remember, on day 1, the player will learn the rules of the game and decide right away which boxes to open. One day later (on day 2), they are allowed to change their decision or stick with their initial decision. \\ \bottomrule
    \end{tabular}
\end{center}

\begin{itemize}
    \item If you want the player to be able to open any number of boxes, up to the maximum of 25, enter '25' into the field below.
    \item If you want to set a limit, enter the maximum number of boxes they should be able to open into the field below. Remember, the player can open anywhere from 0 up to the number you set.
    \item Remember that this game is just one part of a two-day survey that the player will take. They will learn the game rules on day 1. They will always be paid after day 2.
\end{itemize}

Your decision will be implemented for a real player if your response is randomly selected, so take it seriously. It could affect someone else.

\textbf{Maximum number of boxes the player is able to open:}

[Numeric input field, allowable values: 0, 1, …, 25]

\begin{center}
    \begin{tabular}{@{}p{0.125\textwidth}p{0.585\textwidth}p{0.125\textwidth}p{0.125\textwidth}@{}}
        \toprule
        \multicolumn{1}{c}{\textbf{Once}} & \multicolumn{1}{c}{\textbf{EndoDelay}} & \multicolumn{1}{c}{\textbf{Delay}} & \multicolumn{1}{c}{\textbf{Undo}} \\ \midrule
        & Decide when they are to make the choice: \\
        & [Radio button] The player is to make the choice right away (on day 1).\\
        & [Radio button] The player is to make the choice only after one day (on day 2). \\ \bottomrule
    \end{tabular}
\end{center}

\section{Preregistered analyses}

The preregistration may be perused at \url{https://osf.io/m7526}.

All tests are two-sided.

\subsection{Analysis 1}
\label{undopra1}

\noindent\emph{Welch's two-sample $t$-test between} Cap in Once \emph{and} Cap in EndoDelay

\begin{adjustbox}{max width=\textwidth}%
    \begin{tabular}{@{}lcc@{}}
        \toprule
        & Cap in Once & Cap in EndoDelay\\
        \midrule
        Sample size & 639 & 690 \\
        Sample mean & 17.358 & 17.691 \\
        Sample std.dev. & 7.433 & 7.575 \\ \midrule
        Difference in means & \multicolumn{2}{c}{$-0.333$} \\
        $t$ & \multicolumn{2}{c}{$-0.808$} \\
        Degrees of freedom & \multicolumn{2}{c}{1322.557} \\
        $p$-value & \multicolumn{2}{c}{0.419} \\
        95\% confidence interval & \multicolumn{2}{c}{$[-1.141, 0.475]$} \\
        \bottomrule
    \end{tabular}%
\end{adjustbox}

\subsection{Analysis 2}
\label{undopra2}

639 CAs are in Once. 690 are in EndoDelay.

In Once, 434 CAs (67.9\%) set a cap of less than 25. In EndoDelay, 538 CAs (78.0\%) set a cap of less than 25, imposed a waiting period, or both.

Using a test of equal proportions, we find $\chi^2 = 16.6$, $p < 0.001$ (95\% confidence interval for the difference: $[0.051, 0.150]$).

\subsection{Analysis 3}
\label{undopra3}

\noindent\emph{Welch's two-sample $t$-test between} Cap in Once \emph{and} Cap in Delay

\begin{adjustbox}{max width=\textwidth}%
    \begin{tabular}{@{}lcc@{}}
        \toprule
        & Cap in Once & Cap in Delay\\
        \midrule
        Sample size & 639 & 717 \\
        Sample mean & 17.358 & 17.688 \\
        Sample std.dev. & 7.433 & 7.596 \\ \midrule
        Difference in means & \multicolumn{2}{c}{$-0.329$} \\
        $t$ & \multicolumn{2}{c}{$-0.806$} \\
        Degrees of freedom & \multicolumn{2}{c}{1342.249} \\
        $p$-value & \multicolumn{2}{c}{0.421} \\
        95\% confidence interval & \multicolumn{2}{c}{$[-1.131, 0.472]$} \\
        \bottomrule
    \end{tabular}%
\end{adjustbox}

\subsection{Analysis 4}
\label{undopra4}

\noindent\emph{Welch's two-sample $t$-test between} Cap in Delay \emph{and} Cap in Undo

\begin{adjustbox}{max width=\textwidth}%
    \begin{tabular}{@{}lcc@{}}
        \toprule
        & Cap in Delay & Cap in Undo\\
        \midrule
        Sample size & 717 & 656 \\
        Sample mean & 17.688 & 17.291 \\
        Sample std.dev. & 7.596 & 7.784 \\ \midrule
        Difference in means & \multicolumn{2}{c}{$0.396$} \\
        $t$ & \multicolumn{2}{c}{$0.954$} \\
        Degrees of freedom & \multicolumn{2}{c}{1353.618} \\
        $p$-value & \multicolumn{2}{c}{0.340} \\
        95\% confidence interval & \multicolumn{2}{c}{$[-0.419, 1.212]$} \\
        \bottomrule
    \end{tabular}%
\end{adjustbox}

\subsection{Analysis 5}
\label{undopra5}

\noindent\emph{Welch's two-sample $t$-test between} Cap in Undo \emph{and} Cap in Once

\begin{adjustbox}{max width=\textwidth}%
    \begin{tabular}{@{}lcc@{}}
        \toprule
        & Cap in Undo & Cap in Once\\
        \midrule
        Sample size & 656 & 639 \\
        Sample mean & 17.291 & 17.358 \\
        Sample std.dev. & 7.784 & 7.433 \\ \midrule
        Difference in means & \multicolumn{2}{c}{$-0.067$} \\
        $t$ & \multicolumn{2}{c}{$-0.159$} \\
        Degrees of freedom & \multicolumn{2}{c}{1292.491} \\
        $p$-value & \multicolumn{2}{c}{0.874} \\
        95\% confidence interval & \multicolumn{2}{c}{$[-0.897, 0.762]$} \\
        \bottomrule
    \end{tabular}%
\end{adjustbox}

\subsection{Analysis 6}
\label{undopra6}

A Kolmogorov--Smirnov test between the norms in Once and Delay reveals $D$ = 0.041, $p \approx 0.62$. This test is meant as a robustness check, and not intended to reject \citep[14]{undoproposal}.

\subsection{Analysis 7}
\label{undopra7}

\noindent\emph{Welch's two-sample $t$-test between} Error in Once \emph{and} Error in Delay

\begin{adjustbox}{max width=\textwidth}%
    \begin{tabular}{@{}lcc@{}}
        \toprule
        & Error in Once & Error in Delay\\
        \midrule
        Sample size & 628 & 696 \\
        Sample mean & 5.005 & 4.398 \\
        Sample std.dev. & 4.159 & 4.211 \\ \midrule
        Difference in means & \multicolumn{2}{c}{$0.607$} \\
        $t$ & \multicolumn{2}{c}{$2.635$} \\
        Degrees of freedom & \multicolumn{2}{c}{1311.265} \\
        $p$-value & \multicolumn{2}{c}{0.009} \\
        95\% confidence interval & \multicolumn{2}{c}{$[0.155, 1.059]$} \\
        \bottomrule
    \end{tabular}%
\end{adjustbox}

\subsection{Analysis 8}
\label{undopra8}

\noindent\emph{Welch's two-sample $t$-test between} Forecast happiness in Once \emph{and} Forecast happiness in Delay

\begin{adjustbox}{max width=\textwidth}%
    \begin{tabular}{@{}lcc@{}}
        \toprule
        & Forecast happiness in Once & Forecast happiness in Delay\\
        \midrule
        Sample size & 639 & 717 \\
        Sample mean & 4.354 & 4.413 \\
        Sample std.dev. & 1.264 & 1.243 \\ \midrule
        Difference in means & \multicolumn{2}{c}{$-0.059$} \\
        $t$ & \multicolumn{2}{c}{$-0.867$} \\
        Degrees of freedom & \multicolumn{2}{c}{1330.606} \\
        $p$-value & \multicolumn{2}{c}{0.386} \\
        95\% confidence interval & \multicolumn{2}{c}{$[-0.193, 0.075]$} \\
        \bottomrule
    \end{tabular}%
\end{adjustbox}

\subsection{Analyses 9}
\label{undopra9}

\subsubsection{Analysis 9a}

\noindent\emph{Welch's two-sample $t$-test between} Cap in EndoDelay (Chooser in Delay) \emph{and} Cap in Delay

\begin{adjustbox}{max width=\textwidth}%
    \begin{tabular}{@{}lcc@{}}
        \toprule
        & Cap in EndoDelay (Chooser in Delay) & Cap in Delay\\
        \midrule
        Sample size & 267 & 717 \\
        Sample mean & 17.775 & 17.688 \\
        Sample std.dev. & 7.734 & 7.596 \\ \midrule
        Difference in means & \multicolumn{2}{c}{$0.088$} \\
        $t$ & \multicolumn{2}{c}{$0.159$} \\
        Degrees of freedom & \multicolumn{2}{c}{468.939} \\
        $p$-value & \multicolumn{2}{c}{0.874} \\
        95\% confidence interval & \multicolumn{2}{c}{$[-0.997, 1.172]$} \\
        \bottomrule
    \end{tabular}%
\end{adjustbox}

\subsubsection{Analysis 9b}

\noindent\emph{Welch's two-sample $t$-test between} Cap in EndoDelay (Chooser in Once) \emph{and} Cap in Once

\begin{adjustbox}{max width=\textwidth}%
    \begin{tabular}{@{}lcc@{}}
        \toprule
        & Cap in EndoDelay (Chooser in Once) & Cap in Once\\
        \midrule
        Sample size & 423 & 639 \\
        Sample mean & 17.638 & 17.358 \\
        Sample std.dev. & 7.481 & 7.433 \\ \midrule
        Difference in means & \multicolumn{2}{c}{$0.280$} \\
        $t$ & \multicolumn{2}{c}{$0.599$} \\
        Degrees of freedom & \multicolumn{2}{c}{899.650} \\
        $p$-value & \multicolumn{2}{c}{0.550} \\
        95\% confidence interval & \multicolumn{2}{c}{$[-0.638, 1.198]$} \\
        \bottomrule
    \end{tabular}%
\end{adjustbox}

\subsection{Analysis 10}
\label{undopra10}

639 CAs are in Once. 656 are in Undo. 717 are in Delay.

In Once, 205 CAs (32.1\%) did not set a cap (i.e., they set a cap of 25). In Undo, 199 CAs (30.3\%) did not set a cap. In Delay, 253 CAs (35.3\%) did not set a cap.

A 3-sample test for equality of proportions returns $\chi^2 = 3.96$ (2 degrees of freedom), $p = 0.14$.

\subsection{Adjusting for multiple hypothesis testing}
\label{undomht}

\begin{table}[H]
    \begin{adjustbox}{max width=\textwidth}%
        \begin{tabular}{@{}cccccccccccc@{}}
            \toprule
            $p$-value & Subset       & A1    & A2    & A3    & A4    & A5    & A7    & A8    & A9a   & A9b   & A10   \\ \midrule
            Raw       & ---          & 0.419 & 0.000 & 0.421 & 0.340 & 0.874 & 0.009 & 0.386 & 0.874 & 0.550 & 0.138 \\
            Holm adj. & Caps         & 1.000 &       & 1.000 & 1.000 & 1.000 &       &       &       &       &       \\
            Holm adj. & Mechanisms   &       &       &       &       &       & 0.017 & 0.386 &       &       &       \\
            Holm adj. & All          & 1.000 & 0.000 & 1.000 & 1.000 & 1.000 & 0.077 & 1.000 & 1.000 & 1.000 & 1.000 \\ \bottomrule
        \end{tabular}
    \end{adjustbox}
    \caption{Adjusted $p$-values}
    \label{undoadjp}
\end{table}

Table \ref{undoadjp} adjusts $p$-values for multiple hypothesis testing using \citeauthor{holm1979simple}'s (\citeyear{holm1979simple}) method. We group hypotheses into test families by outcome domain \citep[Remark~2.1]{list2023multiple}: the ``Caps'' family (A1, A3--A5) shares the cap as outcome variable; the ``Mechanisms'' family (A7, A8) tests CAs' beliefs about Choosers under deliberation, a distinct outcome domain. Analysis 2 tests substitutability with a distinct outcome variable (proportion intervening) and constitutes a single-test family. Analysis 6 (Section \ref{undopra6} in this Appendix) is omitted because it was not meant to reject. Analyses 9a, 9b, and 10 were preregistered as exploratory and are therefore distinguished from the confirmatory tests \citep{olken2015promises}.

Analysis 2 (Section \ref{undopra2} in this Appendix) relates to hypothesis \ref{undohyp2}. It survives even the most conservative adjustment, where all preregistered tests are jointly adjusted.

Analysis 7 (Section \ref{undopra7} in this Appendix) relates to hypothesis \ref{undohyp6}. It survives the within-family adjustment (adjusted $p = 0.017$). The final row of Table \ref{undoadjp} reports a conservative sensitivity check in which all tests are adjusted jointly, yielding adjusted $p = 0.077$; because this adjustment pools distinct outcome domains, it should be understood as a conservative upper bound. Even under this bound, the $t$-test is the most conservative among common tests. A Wilcoxon rank sum test on the same data reveals $p = 0.00028$ (conservatively adjusted: $p = 0.003$). A two-sample Kolmogorov--Smirnov test shows $p \approx 0.00031$ (conservatively adjusted: $p = 0.003$). Wilcoxon-type tests have been criticized in the literature \citep{divine2018wilcoxon}. We can use Mood's median test to evaluate a null hypothesis of equal medians. When we do so, we find $p \approx 0.005$, conservatively adjusted to $p \approx 0.042$.

Moreover, we can group Undo and Delay, and Once and EndoDelay. Recall from Section \ref{undomentmodresult} that Once and EndoDelay were identical on the experiment's first page, and that Undo and Delay both share deliberation. Note how model 4 in Table \ref{tu1} shows how errors are slightly, though not significantly, lower in EndoDelay than in Once, making the overall grouped error closer to Once and Delay. Yet, after grouping, Welch's $t$-test returns $p = 0.0015$ (conservatively adjusted: $p = 0.013$). Wilcoxon and Kolmogorov--Smirnov tests show, after conservative adjustment, $p < 0.001$. A conservatively adjusted $p < 0.001$ also obtains in a Fisher's exact test on the null hypothesis that the proportion of errors equal to zero (model 1 in Table \ref{tu3} in this Appendix) is equal between the two groups of treatments.

The following table reports conservatively adjusted $p$-values from Welch's $t$-tests, Wilcoxon rank sum tests and Kolmogorov--Smirnov tests with all possible reasonable groupings of treatments:

\begin{adjustbox}{max width=\textwidth}%
    \begin{tabular}[t]{llcccccc}
        \toprule
        Sample 1 & Sample 2 & Mean 1 & Mean 2 & Direction & Welch's $t$ & Wilcoxon & K--S\\
        \midrule
        Once & Delay & 5.005 & 4.398 & $>$ & 0.077 & 0.003 & 0.003\\
        Once & Undo & 5.005 & 4.262 & $>$ & 0.018 & 0.000 & 0.000\\
        Once & Delay \& Undo & 5.005 & 4.333 & $>$ & 0.009 & 0.000 & 0.000\\
        EndoDelay & Delay & 4.715 & 4.398 & $>$ & 1.000 & 0.423 & 1.000\\
        EndoDelay & Undo & 4.715 & 4.262 & $>$ & 0.475 & 0.022 & 0.253\\
        EndoDelay & Delay \& Undo & 4.715 & 4.333 & $>$ & 0.465 & 0.035 & 0.421\\
        Once \& EndoDelay & Delay & 4.854 & 4.398 & $>$ & 0.183 & 0.011 & 0.035\\
        Once \& EndoDelay & Undo & 4.854 & 4.262 & $>$ & 0.039 & 0.000 & 0.001\\
        Once \& EndoDelay & Delay \& Undo & 4.854 & 4.333 & $>$ & 0.013 & 0.000 & 0.000\\
        \bottomrule
    \end{tabular}
\end{adjustbox}%

From this evidence, we conclude that perceived errors appear to be reduced through Chooser deliberation. However, it is a small effect.

\section{Additional tables}
\label{addtable}

\begin{table}[H]
    \begin{center}
        \sisetup{parse-numbers=false, table-text-alignment=center}
        \begin{threeparttable}
            \begin{tabular}{l S[table-format=4.6] S[table-format=4.6]}
                \toprule
                & {Model 1} & {Model 2} \\
                \midrule
                Intercept                   & 0.168^{***}  & 0.072^{***}  \\
                & (0.010)      & (0.020)      \\
                Delay or Undo               & 0.061^{***}  & 0.129^{***}  \\
                & (0.015)      & (0.030)      \\
                Norm                        &              & 0.010^{***}  \\
                &              & (0.002)      \\
                Delay or Undo $\times$ Norm &              & -0.007^{*}   \\
                &              & (0.003)      \\
                \midrule
                Outcome                     & {Error is 0} & {Error is 0} \\
                R$^2$                       & 0.006        & 0.018        \\
                Adj. R$^2$                  & 0.006        & 0.016        \\
                Num. obs.                   &{2649}        &{2649}        \\
                \bottomrule
            \end{tabular}
            \begin{tablenotes}[flushleft]
            \scriptsize{\item $^{***}p<0.001$; $^{**}p<0.01$; $^{*}p<0.05$}
            \end{tablenotes}
        \end{threeparttable}
        \caption{Heterogeneous treatment effects on binary outcome}
        \label{tu3}
    \end{center}
\end{table}

\begin{table}[H]
    \begin{center}
        \sisetup{parse-numbers=false, table-text-alignment=center}
        \begin{adjustbox}{max width=\textwidth}%
            \begin{threeparttable}
                \begin{tabular}{l S[table-format=4.6] S[table-format=4.6] S[table-format=4.6] S[table-format=4.6]}
                    \toprule
                    & {Model 1} & {Model 2} & {Model 3} & {Model 4} \\
                    \midrule
                    Intercept               & 12.158^{***} & 7.950^{***} & 5.005^{***} & 4.471^{***}  \\
                    & (0.225)      & (0.429)     & (0.166)     & (0.405)      \\
                    Undo                    & -1.291^{***} & -1.802^{**} & -0.743^{**} & -1.326^{*}   \\
                    & (0.325)      & (0.602)     & (0.240)     & (0.566)      \\
                    Delay                   & -1.024^{***} & -0.944      & -0.607^{**} & -1.883^{***} \\
                    & (0.306)      & (0.583)     & (0.230)     & (0.547)      \\
                    EndoDelay               & -0.531       & -0.837      & -0.290      & -0.581       \\
                    & (0.315)      & (0.584)     & (0.229)     & (0.541)      \\
                    Norm                    &              & 0.439^{***} &             & 0.055        \\
                    &              & (0.045)     &             & (0.043)      \\
                    Undo $\times$ Norm      &              & 0.069       &             & 0.065        \\
                    &              & (0.062)     &             & (0.058)      \\
                    Delay $\times$ Norm     &              & -0.018      &             & 0.128^{*}    \\
                    &              & (0.061)     &             & (0.058)      \\
                    EndoDelay $\times$ Norm &              & 0.032       &             & 0.030        \\
                    &              & (0.060)     &             & (0.056)      \\
                    \midrule
                    Outcome                 & {Belief}     & {Belief}    & {Error}     & {Error}      \\
                    R$^2$                   & 0.007        & 0.231       & 0.005       & 0.034        \\
                    Adj. R$^2$              & 0.006        & 0.229       & 0.003       & 0.031        \\
                    Num. obs.               &{2661}        &{2649}       &{2649}       &{2649}        \\
                    \bottomrule
                \end{tabular}
                \begin{tablenotes}[flushleft]
                \scriptsize{\item $^{***}p<0.001$; $^{**}p<0.01$; $^{*}p<0.05$}
                \end{tablenotes}
            \end{threeparttable}
        \end{adjustbox}
        \caption{Heterogeneous treatment effects (without treatment grouping)}
        \label{tu4}
    \end{center}
\end{table}

\begin{table}[H]
    \begin{center}
        \sisetup{parse-numbers=false, table-text-alignment=center}
        \begin{threeparttable}
            \begin{tabular}{l S[table-format=4.6] S[table-format=4.6]}
                \toprule
                & {Model 1} & {Model 2} \\
                \midrule
                Intercept               & 0.161^{***}  & 0.062^{*}    \\
                & (0.015)      & (0.028)      \\
                Undo                    & 0.084^{***}  & 0.164^{***}  \\
                & (0.022)      & (0.044)      \\
                Delay                   & 0.055^{*}    & 0.112^{**}   \\
                & (0.021)      & (0.041)      \\
                EndoDelay               & 0.014        & 0.018        \\
                & (0.021)      & (0.040)      \\
                Norm                    &              & 0.010^{***}  \\
                &              & (0.003)      \\
                Undo $\times$ Norm      &              & -0.008       \\
                &              & (0.004)      \\
                Delay $\times$ Norm     &              & -0.006       \\
                &              & (0.004)      \\
                EndoDelay $\times$ Norm &              & -0.000       \\
                &              & (0.004)      \\
                \midrule
                Outcome                 & {Error is 0} & {Error is 0} \\
                R$^2$                   & 0.007        & 0.019        \\
                Adj. R$^2$              & 0.006        & 0.016        \\
                Num. obs.               &{2649}        &{2649}        \\
                \bottomrule
            \end{tabular}
            \begin{tablenotes}[flushleft]
            \scriptsize{\item $^{***}p<0.001$; $^{**}p<0.01$; $^{*}p<0.05$}
            \end{tablenotes}
        \end{threeparttable}
        \caption{Heterogeneous treatment effects on binary outcome (without treatment grouping)}
        \label{tu5}
    \end{center}
\end{table}

\section{Chooser experiment protocol}
\label{chooserprotocol}

The Chooser experiment was programmed in oTree \citep{chen2016otree} and conducted as an online experiment at the Cologne Laboratory for Economic Research in October 2024. Four Choosers participated (one per treatment). For the EndoDelay treatment, the randomly selected CA chose not to impose the waiting period; consequently, the EndoDelay Chooser faced the same protocol as the Once Chooser.

\subsection{Two-day structure}

The experiment spanned two days (Wednesday, October 23, 2024, and Thursday, October 24, 2024). After completing the Day~1 pages, participants reached a password-gated page; the password was provided on Day~2 and was required to continue.
The temporal structure of each treatment was as follows:
\begin{itemize}[nosep]
    \item \textbf{Once:} Day~1: consent, instructions, BRET decision. Day~2: CRT, end.
    \item \textbf{Delay:} Day~1: consent, instructions, CRT. Day~2: BRET decision, end.
    \item \textbf{Undo:} Day~1: consent, instructions, BRET decision~1. Day~2: BRET decision~2 (optional revision), CRT, end.
\end{itemize}
In all treatments, the Cognitive Reflection Test \citep{frederick2005cognitive} was administered as a filler task. Performance on the CRT did not affect payment.

\subsection{BRET interface}

Choosers were presented with a $5 \times 5$ visual grid of 25 clickable boxes. Boxes could be selected individually by clicking or collectively via a slider control that randomly assigned selections. Selected boxes were visually highlighted.

The cap set by the randomly selected CA was communicated to the Chooser in the instructions; in the Delay implementation, further selections beyond the cap were prevented by the interface. In the Undo treatment, the Day~2 revision page pre-loaded the Chooser's Day~1 selection, allowing changes from that starting point.

\subsection{Determination of the curveball}

The position of the curveball was determined by the first number drawn in the Eurojackpot lottery on Friday of the same week. Numbers 1--25 mapped directly to box positions; numbers 26--50 were mapped by subtracting~25. This procedure assigns a uniform $1/25$ probability to each box and is publicly verifiable. Choosers were informed of this mechanism in the instructions and told that the Eurojackpot draw video (hosted on YouTube and the official Eurojackpot website) would serve as proof of the outcome.

\subsection{Payment}

Choosers received a \texteuro1 show-up fee for participating on both days. BRET earnings were stated in U.S.\ dollars (\$20 per opened box, unless the curveball was among the opened boxes) and converted to euros using the official ECB exchange rate published on the day of the draw. Payments were made via SEPA bank transfer.

\end{document}